\documentstyle[12pt]{article}

\begin{document}
\
\vspace{1cm}

\begin{centering}
{\large

Vortex tubes in velocity fields of laboratory isotropic turbulence

}
\vspace{2cm}

Hideaki Mouri       \footnote{Corresponding author. E-mail: hmouri@mri-jma.go.jp},
Akihiro Hori        \footnote{Affiliated with Japan Weather Association},
Yoshihide Kawashima \footnote{Affiliated with Tsukuba Technology Research}

Meteorological Research Institute, Nagamine 1-1, Tsukuba 305-0052, Japan 

\vspace{3cm}

To appear in Physics Letters A

\vspace{2cm}
\end{centering}

\begin{abstract}
The streamwise and transverse velocities are measured simultaneously in grid turbulence at Reynolds numbers 100--300. We extract typical intermittency patterns, which are consistent with velocity profiles of Burgers and Lamb--Oseen vortices. The radii of these vortex tubes are estimated to be several of the Kolmogorov length.
\end{abstract}

\newpage
\small
\section{Introduction}

By using direct numerical simulations \cite{j93} and bubble-visualization experiments \cite{d91}, it has been established that turbulence contains vortex tubes. Regions of intense vorticity are organized into tubes. Their radii and lengths are, respectively, of the orders of the Kolmogorov length $\eta$ and the integral length $L$. Their lifetimes are several turnover times of the largest eddies. Vortex tubes occupy a small fraction of the volume, and are embedded in a background flow which is random and of large scales (see also reviews \cite{f95,sa97,s99}).

The most familiar model for the tubes is the Burgers vortex \cite{b48}. This is an axisymmetric flow in a strain field. In cylindrical coordinates, they are written as
\begin{equation}
\label{a}
  u_{\theta} \propto \frac{2 \nu}{a r} 
             \left( 1 - \exp \left( - \frac{a r^2}{4 \nu} \right) \right)
  \quad (a > 0),
\end{equation}
and
\begin{equation}
\label{b}
\left( u_r , u_{\theta}, u_z \right) = 
\left( - \frac{1}{2} a r, 0, a z\right).
\end{equation}
Here $\nu$ is the kinetic viscosity. The above equation (\ref{a}) describes a rigid-body rotation for small radii, and a circulation decaying in radius for large radii. The velocity is maximal at $r$ = $r_0$ = $2.24 (\nu / a)^{\frac{1}{2}}$. Thus $r_0$ is regarded as the tube radius. The same circulation pattern is expected for the Lamb--Oseen vortex, i.e., an unstretched diffusing tube \cite{l32}.

The effect of vortex tubes on the velocity field is of great interest. This is especially the case at high Reynolds numbers, $Re_{\lambda}$ $\gg$ 100. They are achieved only in experiments, where a measurement is made with a probe suspended in the flow, and merely a one-dimensional cut of the velocity field is obtained. The velocity signal at small scales is enhanced in a fraction of the volume \cite{bt49}. This small-scale intermittency is attributable to vortex tubes. However, measurements often deal with the velocity component in the mean-flow direction alone (hereafter, the streamwise velocity $u$). The transverse velocity $v$ is more suited to detecting rotational flows such as those associated with vortex tubes \cite{n97,cg99,mtk99}.

This Letter reports simultaneous measurements of the streamwise and transverse velocities in high-Reynolds-number flows behind a grid. Grid turbulence is almost isotropic and free from external influences such as mean shear, and is hence suited to studying universal properties of turbulence \cite{bt49,t92}. With a conditional averaging technique, we extract typical intermittency patterns, which turn out to agree with velocity profiles of Burgers and Lamb--Oseen vortices. Then energetic importance of the vortex tubes is studied as a function of the scale.

\section{Experiments}

The experiments were done in two wind tunnels of Meteorological Research Institute. Their test sections are of 3 $\times$ 2 $\times$ 18 and 0.8 $\times$ 0.8 $\times$ 3 m in size. Turbulence was produced by placing a grid across the entrance to the test section. The grid consists of two layers of uniformly spaced rods, the axes of which are perpendicular to each other. We used three grids. The separation of the axes of their adjacent rods are $M$ = 10, 20, and 40 cm. The cross sections of the rods are, respectively, 2 $\times$ 2, 4 $\times$ 4, and 6 $\times$ 6 cm. The mean wind was set to be $U$ $\simeq$ 4, 8, or 12 m~s$^{-1}$. We consequently conducted 8 sets of experiments as summarized in Table 1.


The streamwise ($U + u$) and transverse ($v$) velocities were measured simultaneously with a hot-wire anemometer. The anemometer is composed of a crossed-wire probe and a constant temperature system. The probe was positioned on the tunnel axis at $d$ = 2, 3, or 6 m downstream of the grid. The hot wires are 5 $\mu$m in diameter, 1.25 mm in effective length, 1.25 mm in separation, and oriented at $\pm 45 ^{\circ}$ to the mean-flow direction. Pitch-angle calibration was done before each of the measurements. The signal was low-pass filtered at $f_c$ = 4--12 kHz with 24 dB/octave, and sampled digitally at $f_s$ = 8--24 kHz with 16-bit resolution. To avoid aliasing, the sampling frequency was set to be twice of the filter cutoff frequency. The entire length of the signal was as long as 2 $\times$ 10$^7$ points, since we have to minimize statistical uncertainties.

The turbulence level, i.e., the ratio of the root-mean-square value of the streamwise fluctuation $\langle u^2 \rangle ^{\frac{1}{2}}$ to the mean streamwise velocity $U$, was always less than 10\% (Table 1). This good characteristic of grid turbulence allows us to rely on the frozen-eddy hypothesis of Taylor, $\partial / \partial t $ = $-U \partial / \partial x$, which converts temporal variations into spatial variations in the mean-wind direction \cite{t38}.

To obtain the maximum spatial resolution, we set $f_c$ in such a way that $U / f_c$ is comparable to the probe size, $\sim 1$ mm. Since the probe is larger than the Kolmogorov length (see below), the smallest-scale motions of the flow were filtered out. The present spatial resolution is nevertheless typical of hot-wire anemometry \cite{cg99,d97}. Though the RELIEF technique achieves a greater resolution \cite{n97}, it is applicable only to the transverse velocity.

The flow parameters such as the integral length $L$, the Kolmogorov length $\eta$, the Taylor microscale $\lambda$, and the Reynolds number $Re_{\lambda}$ are given in Table 1. The $Re_{\lambda}$ values range from 100 to 300. They exceed those in almost all the past studies of grid turbulence \cite{sa97,bt49,t92}. This is because our wind tunnels as well as our grids are large. The present data provide for the first time an opportunity to study vortex tubes in isotropic turbulence at high Reynolds numbers.

\section{Conditional average}

To examine the presence or absence of vortex tubes, we extract typical intermittency patterns in the streamwise ($u$) and transverse velocities ($v$). The patterns are obtained by averaging the signals centered at the position where the transverse-velocity increment $\delta v(s)$ = $v(x+s) - v(x)$ is enhanced over its root-mean-square value by more than a factor of 3, i.e., $\vert \delta v(s) \vert$ $>$ $3 \langle \delta v(s)^2 \rangle ^{\frac{1}{2}}$. Here the scale $s$ is set to be the spatial resolution $U / f_c$. Such an event is detected 2--5 times per the integral length. When the increment is negative, we inverse the sign of the $v$ signal. The results for experiments 1, 3, 7, and 8 are shown in Fig. 1 (solid lines). We applied the conditional averaging also for the other experiments and with other threshold values. Their results are consistent with those in Fig. 1.


For reference, we show velocity profiles of Burgers vortices (dotted lines). The transverse component is from the circulation flow (\ref{a}), while the streamwise component is from the radial inflow $- \frac{1}{2} a r$ of the strain field (\ref{b}). Here it is assumed that the tube center passes through the probe position and the tube axis is perpendicular to the streamwise and transverse directions. The tube radius $r_0$ was determined so as to reproduce the $v$ pattern for each of the experiments.

Grid turbulence exhibits nearly the same $v$ pattern as that of a Burgers vortex. Tubes are surely ubiquitous at $Re_{\lambda}$ $\simeq$ 100--300. Their radii are estimated to be several of the Kolmogorov length ($r_0$ $\simeq$ 6--$8 \eta$). Since these estimates are well above the $s$ values, they should represent typical radii of vortex tubes. Similar tube radii have been obtained from direct numerical simulations \cite{j93}. 

The $u$ pattern of grid turbulence exhibits a trace of the strain field. Thus at least some of the vortex tubes are of Burgers type. However, the observed strain is less significant than those of Burgers vortices. There could exist unstretched tubes such as Lamb--Oseen vortices. In numerical turbulence \cite{j93}, vortex tubes are not always oriented to the stretching direction. Once a tube is stretched to a high amplitude, it decouples from the original strain field. This could be the case in grid turbulence \cite{t92}.

We applied the conditional averaging to larger values of $s$ (not shown here). The velocity patterns for $s < L$ have nearly the same shapes as in Fig. 1. However, with increasing the scale $s$, the corresponding tube radius $r_0$ becomes large, while the amplitude of the $u$ pattern becomes small as compared with that of the $v$ pattern. Velocity signals at large scales reflect tubes which are inclined to the mean-wind direction. The inclination does not change the shape of the velocity pattern, but the tube radius along the mean-wind direction $r_{mw}$ is greater than its true radius \cite{mtk99,b96}. In addition, there could exist large-radius unstretched tubes which are not coupled with the strain field. 


Velocity patterns are extracted also for enhancements of the streamwise-velocity increment $\delta u(s)$ = $u(x+ s) - u(x)$. The results for $\delta u(s)$ $>$ $+3 \langle \delta u(s)^2 \rangle ^{\frac{1}{2}}$ are shown in Fig. 2, while those for $\delta u(s)$ $<$ $-3 \langle \delta u(s)^2 \rangle ^{\frac{1}{2}}$ are shown in Fig. 3. The $v$ pattern, which was obtained by conditioning over the sign of the local slope, is close to a velocity profile of a Burgers or Lamb--Oseen vortex. Thus an enhancement of $\delta u(s)$ is caused by a vortex tube. The $u$ pattern is explained by the circulation flow (\ref{a}) crossing the probe at some distance \cite{b96}, together with the strain field (\ref{b}). The event $\delta u(s)$ $<$ $-3 \langle \delta u(s)^2 \rangle ^{\frac{1}{2}}$ is detected two times more frequently than the event $\delta u(s)$ $>$ $+3 \langle \delta u(s)^2 \rangle ^{\frac{1}{2}}$. This fact is consistent with the $u$ pattern in Fig. 1, which exhibits a negative slope at the tube position.

We comment on controversial results in Figs. 1--3. First, the $v$ pattern is somewhat extended as compared with the velocity profile of a Burgers vortex, especially in experiments 7 and 8, where the Reynolds number is high. Vortex tubes in grid turbulence might not be exactly the same as Burgers or Lamb--Oseen vortices. There might exist additional contributions from inclined tubes. Second, the $u$ pattern is embedded in a one-sided excursion, $u$ $\simeq$ a few tenths of $\langle u^2 \rangle ^{\frac{1}{2}}$ $>$ 0, which extends to $x$ $\simeq$ $\pm L$. This might be due to an artifact of the frozen-eddy hypothesis. The streamwise velocity has a fluctuation of the order of $\langle u^2 \rangle ^{\frac{1}{2}}$ over the scale $L$. Thus the speed at which a vortex tube crosses the probe is higher or lower than the mean wind speed. The observed radius of the tube is accordingly smaller or larger than its $r_{mw}$ value. Since the scale $s$ is at around the typical radius of vortex tubes, inclined tubes with $r_{mw}$ $>$ $s$ are more numerous than tubes with $r_{mw}$ $<$ $s$. Fast-moving tubes with $r_{mw}$ $>$ $s$ could be captured in our averaging process for the scale $s$ and cause the observed $u$ excursion. We admit that these interpretations are tentative. It is important to conduct detailed analyses of data obtained under wider ranges of conditions.

\section{Flatness factor}

To study energetic importance of vortex tubes as a function of the scale, we compute the flatness factor of the transverse-velocity increment $\delta v(s)$ = $v(x+s)-v(x)$:
\begin{equation}
   {\rm flatness\ factor} = \frac{ \langle \delta v(s)^4 \rangle }
                                 { \langle \delta v(s)^2 \rangle ^2}.
\end{equation}
The flatness factor serves as a measure of the relative importance between vortex tubes and the background flow at each of the scales. If the velocity increments follow a Gaussian distribution, the flatness factor is equal to 3. If the tail of the probability distribution of the velocity increments is more pronounced than that of a Gaussian, the flatness factor is higher than 3. The velocity-increment distribution observed at $s$ $\ge$ $L$, where the background flow is predominant, is close to a Gaussian \cite{f95,s99,n97}. On the other hand, the $v$ pattern for an enhancement of $\delta v(s)$ at $s$ $<$ $L$ is close to a velocity profile of a Burgers or Lamb--Oseen vortex, i.e., a model for the vortex tubes (\S3).


Fig. 4 shows the flatness factor for the transverse velocity $v$ in experiments 1, 3, 7, and 8. The abscissa is the scale $s$ normalized by the Kolmogorov length $\eta$. We indicate the integral length $L$ and the Taylor microscale $\lambda$ (arrows). At large scales, the flatness factor is equal to the Gaussian value of 3. This is because the velocity increments reflect the background flow. As the scale is decreased from the integral length, the flatness factor begins to increase. This is due to a progressive contribution of vortex tubes. The tubes affect velocity increments at scales around their observed sizes, which could be as large as their lengths ($\simeq L$). Around the Taylor microscale, the increase becomes significant.

Fig. 4 also shows the flatness factor for the streamwise velocity $u$, which is less enhanced than that for the transverse velocity $v$. This finding is consistent with past experimental and numerical studies \cite{j93,d97}. From our results in Figs. 2 and 3, the streamwise flatness factor is expected to reflect the energetic importance of vortex tubes, as in the case of the transverse flatness factor. However, the $u$ pattern for a vortex tube is less pronounced than the $v$ pattern. The circulation flow of a vortex tube contributes to the streamwise velocity only if its axis passes at some distance from the probe. Moreover, the streamwise-velocity increment suffers from the strain field.


The flatness factors in Fig. 4 are displayed in order of increasing the Reynolds number (from top to bottom). There appears to exist a trend of increasing the small-scale flatness. This appearance is confirmed in Fig. 5, which shows the flatness factors at the fixed scales $s$ = $10 \eta$ (a) and $\lambda$ (b). Those for the streamwise and transverse velocities increase in a similar manner with the $Re_{\lambda}$ value. In both the velocity components, vortex tubes are increasingly predominant over the background flow at a higher Reynolds number. The increase of the flatness factors is more significant at the smaller scale $s$ = $10 \eta$, where the background flow is less important. We also studied the flatness factors at $s$ $\simeq$ $L/2$ (not shown here), which appear to depend on the grid size $M$ rather than the $Re_{\lambda}$ value.

\section{Conclusion}

To summarize, vortex tubes are important to the intermittency pattern of the transverse velocity at $Re_{\lambda}$ $\simeq$ 100--300. The pattern is well approximated as a velocity profile of a Burgers or Lamb--Oseen vortex, if its radius is several of the Kolmogorov length (Fig. 1). The strain field associated with the tube is present in the streamwise velocity. With increasing the Reynolds number, vortex tubes become more predominant over the background flow (Figs. 4 and 5). We believe that these findings are new and convincing, owing to long records of both the streamwise and transverse velocities in isotropic turbulence at high Reynolds numbers. Similar experiments of grid turbulence at much higher Reynolds numbers, $Re_{\lambda}$ $\gg$ 300, would be crucial to proceed further \cite{sa97,s99}.

\vspace{0.5cm}
This work was financially supported in part as `basic atomic energy research' by the Science and Technology Agency, Japan. The authors are grateful to the referee for useful comments and M. Takaoka for interesting discussion.



\newpage
Table 1. 
Summary of experimental conditions and turbulence characteristics: grid size $M$, distance from the grid $d$, sampling frequency $f_s$ = $2 f_c$, mean wind speed $U$, root-mean-square values of the streamwise and transverse fluctuations $\langle u^2 \rangle ^{\frac{1}{2}}$ and $\langle v^2 \rangle ^{\frac{1}{2}}$, kinematic viscosity $\nu$, mean energy dissipation rate $\langle \varepsilon \rangle$ = $15 \nu \langle (\partial u/ \partial x )^2 \rangle$, integral length $L$ = $\int \langle u(x+s) u(x) \rangle / \langle u^2 \rangle {\rm d} s$, Taylor microscale $\lambda$ = $( \langle u^2 \rangle / \langle (\partial u/ \partial x )^2 \rangle )^{\frac{1}{2}}$, Kolmogorov length $\eta$ = $(\nu ^3 / \langle \varepsilon \rangle )^{\frac{1}{4}}$, and Reynolds number $Re_{\lambda}$ = $\langle u^2 \rangle ^{\frac{1}{2}} \lambda / \nu$. Velocity derivatives were obtained as $\partial u / \partial x$ = $( 8 u(x+ {\mit\Delta} )- 8 u(x- {\mit\Delta} )-u(x+ 2 {\mit\Delta} )+u(x- 2 {\mit\Delta} ))/ 12 {\mit\Delta}$, where $\mit\Delta$ is the sampling interval $U/f_s$. The experiments 1 and 2 were made in the 0.8 $\times$ 0.8 $\times$ 3 m tunnel, while the other experiments were made in the 3 $\times$ 2 $\times$ 18 m tunnel. To check the isotropy, the transverse velocity was also used to compute the $L$, $\lambda$, and $\eta$ values, via the relations $L$ = $2 \int \langle v(x+s) v(x) \rangle / \langle v^2 \rangle {\rm d} s$ and $\langle (\partial u/ \partial x )^2 \rangle$ = $\langle (\partial v/ \partial x )^2 \rangle /2$. The $L$ value derived from the $v$ data is smaller by a factor of 0.5--0.6 than that derived from the $u$ data. Thus our grid turbulence is anisotropic at largest scales. On the other hand, the $\lambda$ and $\eta$ values derived from the $v$ data agree with those derived from the $u$ data within $\le$ 1\%. The exceptions are the results for experiments 1 and 2, where the discrepancy is 10--20\%. This anisotropic character is consistent with the fact that their measured ratios of $\langle u^2 \rangle ^{\frac{1}{2}} / \langle v^2 \rangle ^{\frac{1}{2}}$ are not very close to unity.

\vspace{10mm}
\begin{center}

\begin{tabular}{ccccccccccccc}
\hline
exp. & 
$M$ &
$d$ &
$f_s$ &
$U$ & 
$\langle u^2 \rangle ^{\frac{1}{2}}$ &
$\langle v^2 \rangle ^{\frac{1}{2}}$ &
$\nu$ &
$\langle \varepsilon \rangle$ &
$L$ &
$\lambda$ &
$\eta$ &
$Re_{\lambda}$ \\
  &
cm &
m &
kHz &
m s$^{-1}$ &
m s$^{-1}$ &
m s$^{-1}$ &
cm$^2$ s$^{-1}$ &
m$^2$ s$^{-3}$ &
cm &
cm &
cm & \\
\hline
1 & 10 & 2 & 8  & 4.76 & 0.234 & 0.193 & 0.145 & 0.282  & 9.75 & 0.650 & 0.0322 & 105 \\
2 & 10 & 2 & 16 & 9.60 & 0.475 & 0.406 & 0.145 & 2.01   & 9.97 & 0.494 & 0.0197 & 162 \\
3 & 20 & 3 & 8  & 4.27 & 0.257 & 0.255 & 0.147 & 0.182  & 16.1 & 0.897 & 0.0364 & 157 \\
4 & 20 & 3 & 16 & 8.38 & 0.506 & 0.500 & 0.147 & 1.45   & 17.1 & 0.624 & 0.0216 & 215 \\
5 & 20 & 3 & 24 & 12.7 & 0.767 & 0.761 & 0.149 & 4.90   & 17.6 & 0.518 & 0.0161 & 267 \\
6 & 40 & 6 & 8  & 4.39 & 0.220 & 0.213 & 0.147 & 0.0718 & 17.2 & 1.22  & 0.0459 & 182 \\
7 & 40 & 6 & 16 & 8.70 & 0.446 & 0.427 & 0.147 & 0.595  & 17.2 & 0.858 & 0.0270 & 260 \\
8 & 40 & 6 & 24 & 13.0 & 0.665 & 0.642 & 0.149 & 2.09   & 15.6 & 0.688 & 0.0199 & 307 \\
\hline
\end{tabular}
\end{center}


\newpage

\section*{Figure captions}
\begin{description}

\item[Fig. 1] Conditional averages of the streamwise ($u$) and transverse ($v$) velocities for $\vert \delta v(s) \vert$ $>$ $3 \langle \delta v(s)^2 \rangle ^{\frac{1}{2}}$ with $s$ = $U / f_c$. The abscissa is the spatial position $x$ normalized by the Kolmogorov length $\eta$. The velocities are normalized by the $\langle v^2 \rangle ^{\frac{1}{2}}$ value. For experiments 1, 3, 7, 8, respectively, the peak amplitudes correspond to 0.179, 0.198, 0.276, and 0.443 m~s$^{-1}$. The values of $3 \langle \delta v(s)^2 \rangle ^{\frac{1}{2}} / \langle v^2 \rangle ^{\frac{1}{2}}$ are 0.778, 0.498, 0.527, and 0.647. The detection rates per the integral length are 1.99, 4.00, 4.51, and 4.00. The profiles of Burgers vortices are shown with dotted lines. Their radii $r_0$ are $5.54 \eta$, $7.35 \eta$, $6.03 \eta$, and $8.15 \eta$. 

\item[Fig. 2] Conditional averages of the streamwise ($u$) and transverse ($v$) velocities for $\delta u(s)$ $>$ $+3 \langle \delta u(s)^2 \rangle ^{\frac{1}{2}}$ with $s$ = $U / f_c$. The abscissa is the spatial position $x$ normalized by the Kolmogorov length $\eta$. The velocities are normalized by the $\langle u^2 \rangle ^{\frac{1}{2}}$ value. For experiments 1, 3, 7, 8, respectively, the peak amplitudes correspond to 0.256, 0.260, 0.354, and 0.538 m~s$^{-1}$. The values of $3 \langle \delta u(s)^2 \rangle ^{\frac{1}{2}} / \langle u^2 \rangle ^{\frac{1}{2}}$ are 0.529, 0.346, 0.359, and 0.445. The detection rates per the integral length are 0.464, 0.938, 1.16, and 1.16. 

\item[Fig. 3] Conditional averages of the streamwise ($u$) and transverse ($v$) velocities for $\delta u(s)$ $<$ $-3 \langle \delta u(s)^2 \rangle ^{\frac{1}{2}}$ with $s$ = $U / f_c$. The abscissa is the spatial position $x$ normalized by the Kolmogorov length $\eta$. The velocities are normalized by the $\langle u^2 \rangle ^{\frac{1}{2}}$ value. For experiments 1, 3, 7, 8, respectively, the peak amplitudes correspond to 0.246, 0.248, 0.338, and 0.518 m~s$^{-1}$. The detection rates per the integral length are 1.08, 2.18, 2.37, and 2.04. 

\item[Fig. 4] Flatness factor of the increments of the streamwise ($u$) and transverse ($v$) velocities. The abscissa is the scale $s$ normalized by the Kolmogorov length $\eta$. Arrows denote the integral length $L$ and the Taylor microscale $\lambda$. Dotted lines denote the Gaussian value of 3.

\item[Fig. 5] Flatness factor of the velocity increments at $s$ = $10 \eta$ (a) and $s$ = $\lambda$ (b). The abscissa is the Reynolds number $Re_{\lambda}$. Open circles denote the streamwise velocity while filled circles denote the transverse velocity. We also show the experiment numbers. The scale $s$ = $10 \eta$ is, roughly speaking, the minimum scale resolved in all the experiments.

\end{description}
\end{document}